\documentclass[preprintnumbers,amsmath,amssymb]{revtex4}

\usepackage{graphicx}
\usepackage{bm}
\newtheorem{corollary}{Corollary}[section]
\newtheorem{remark}{Remark}[section]
\newtheorem{definition}{Definition}[section]
\newtheorem{theorem}{Theorem}[section]
\newtheorem{proposition}{Proposition}[section]
\newtheorem{lemma}{Lemma}[section]

\begin{document}

\title{Exponentially stable breather solutions in
nonautonomous   dissipative  nonlinear Schr\"odinger lattices}

\author{Dirk Hennig}
\email[Email: ]{dirkhennig@uth.gr}
\affiliation{Department of Mathematics, University of Thessaly, Lamia GR35100, Greece}

\date{\today}

\begin{abstract}
\noindent  We consider  damped and forced discrete nonlinear Schr\"odinger equations on 
the lattice $\mathbb{Z}$. First we establish the existence of periodic and quasiperiodic 
breather solutions for periodic and quasiperiodic driving, respectively. Notably, quasiperiodic breathers 
cannot exist in the system without damping and driving. 
Afterwards the existence of a global uniform attractor for the dissipative dynamics of the system is shown. For strong dissipation we prove that the   global uniform attractor  has finite 
fractal dimension and consists of a single trajectory that is  confined to a finite 
dimensional subspace of the infinite dimensional phase space, attracting any bounded set 
in phase space exponentially fast. Conclusively, for strong damping and periodic (quasiperiodic) forcing the single periodic (quasiperiodic) breather solution possesses a finite number of modes and is exponentially stable.
\end{abstract}

\maketitle

\section{Introduction}
We study the dynamics of the 
following  damped and driven general discrete nonlinear Schr\"odinger equation (ddDNLS) on the 
infinite one-dimensional lattice
\begin{equation}
i\frac{d \psi_n}{dt}+\kappa\,[\psi_{n+1}-2\psi_n+\psi_{n-1}]+
\nu F(|\psi_n|^2)\psi_n=-i\gamma\psi_n+g_n,\,\,\,n \in \mathbb{Z},\,\,t>t_0,\,\,t_0\in \mathbb{R}, \label{eq:system}
\end{equation}
supplemented with the initial data 
\begin{equation}
 \psi_n(t_0)=\psi_{t_0,n},\,\,\, n\in \mathbb{Z},\label{eq:icsystem}
\end{equation}
with  $\psi_n \in {\mathbb{C}}$.  The parameter  $\gamma>0$ determines the strength of  damping. We assume $\nu,\kappa >0$ so that (\ref{eq:system}) is of focusing type.

\begin{remark}
The positive parameter $\nu$ in (\ref{eq:system}) can be absorbed via the scaling $x \mapsto \sqrt \nu x$, thus without loss of generality we can take $\nu = 1$.
\end{remark}
 
\vspace*{0.5cm}
$(g_n(t))_{n\in \mathbb{Z}}=g(t)$, is an external spatio-temporal driving field. $F(|\psi|^2)\psi$ is a general nonlinear term on which 
we assume the following condition:


\noindent {\bf{H1:}} 
$F\in C({\mathbb{R}}_+,{\mathbb{R}})$ for ${\mathbb{R}}_+=[0,\infty)$, $F(0)=0$. 
There are constants $a>0$, $b>0$ such that 
\begin{eqnarray}
 |F(|\psi|^2)\psi-F(|\phi|^2)\phi|&\le& a(|\psi|^b+|\phi|^b)|\psi-\phi|.\label{eq:A2}
\end{eqnarray}

The class of nonlinearities $F(x)$ obeying {\bf{H1}} includes the following important cases: 
\\[2mm]
{(i) The general power nonlinearity $F(x) = x^p$.\\
{(ii) The saturable nonlinearity $F(x) = -\dfrac{1}{1+x}$.}

\vspace*{0.5cm}

System (\ref{eq:system}) with a time-independent force was studied in \cite{Nikos},\cite{Dirk}. When $\gamma=0$ and $g=0$ in (\ref{equation:system}), a  discrete nonlinear Schr\"odinger equation (DNLS) 
with general nonlinearity $F(x)$ results, which is time-reversible and possesses a  Hamiltonian structure.
With view to applications in various fields of physics and biophysics, 
the DNLS represents 
an important lattice system  that may exhibit coherent structures in the form of spatially localised (soliton-like) solutions and breather solutions.
We refer  to \cite{DNLS},\cite{Kevrekidis} concerning  discrete nonlinear Schr\"odinger equations and their applications. 
Indeed,  provided the coupling strength $\kappa$ remains sufficiently small, the existence of a family of breathers for the standard (cubic) DNLS, $F(x)=x^2$, is well established \cite{MacKay},\cite{Aubry},\cite{Weinstein}. These breathers are stationary solutions with corresponding functions $\psi_n(t)=\phi_n\exp(-i \omega t)$ representing  standing localised waves on the lattice. 
The trivial phase solution, $\psi_n(t)=\phi_n\exp(-i \omega t)$, is not the most general form of stationary solutions of the standard DNLS. We underline that for the standard DNLS its continuous band of linear frequencies rules out the existence of quasiperiodic solutions because resonances with the continuous band of frequencies cannot be excluded.
The DNLS breathers are stationary states of the DNLS for which 
most of the excitation energy remains stored in (or about) a single site. 
For nonzero dissipation, $\gamma>0$, the picture changes drastically because these breather solutions get destroyed and their excitation energy will be dissipated in the course of time
In order that the damped system exhibits coherent structures, such as breathers, at all, some (external) forcing is required compensating the losses caused by the effect of dissipation.

In fact, in the present paper we prove in Section \ref{section:localised} that in the presence 
of damping there exist periodic and quasiperiodic breather solutions under periodic respectively 
quasiperiodic drivings, respectively. In contrast to the Hamiltonian (time-reversible) case,   where there exists a family of breather solutions 
(lying on infinite-dimensional tori in the infinite dimensional phase space), 
we prove that for the strong damping there exists a single periodic and quasiperiodic breather solution, respectively, that  attracts any bounded set in phase space exponentially fast. Moreover, the single solution  is confined to a finite-dimensional subspace of the infinite dimensional phase space
and coincides with the global uniform attractor for the dissipative dynamics of (\ref{eq:system}).

Our  main results for the strong damping regime are summarised in the following theorem:

\begin{theorem}[Single solution attracting any bounded set of $l^2$ exponentially]\label{theorem:uniquesolution}

Consider the ddDNLS  (\ref{eq:system}). 
Let
\begin{equation}
\gamma>\left(a||g||_{C_b(\mathbb{R},l^2)}^b\right)^{1/(1+b)}.\label{eq:gammaunique}
\end{equation}

Then the system (\ref{eq:system}) possesses a global uniform attractor that consist of a single  solution attracting any bounded set of $l^2$ exponentially fast. 
The global uniform attractor has finite fractal dimension, so that the dynamics of system (\ref{eq:system}) is confined to a finite-dimensional subspace of $l^2$.
\end{theorem}

We prove Theorem  \ref{theorem:uniquesolution} in Sections \ref{section:attractors}-\ref{section:strong} utilising methods from the modern theory of infinite-dimensional dynamical systems \cite{Hale}-\cite{Chueshov}. In general, 
the dissipative dynamics for  infinite lattice dynamical systems (LDS) has attracted considerable interest recently \cite{Babin}-\cite{Han}.
As a first step, in the next section  we prove the existence of a global unique solution for the initial value problem (IVP) (\ref{eq:system}),(\ref{eq:icsystem}. For driving forces contained in the space of bounded functions on the real line we assign a family of processes to the nonautonomous DLS. 
To demonstrate the existence of a uniform attractor for the system  we introduce  a semigroup of nonlinear operators for this family of  processes  in an extended phase space. Afterwards, using the semigroup theory, we prove that the semigroup is asymptotically compact and point dissipative, and hence, has a global attractor from  which one infers on the existence of a uniform attractor for the processes. We stress that our method utilised for the proof of asymptotic tail end property 
differs from the standard method used in the literature \cite{Babin}-\cite{Han}. 

After having established the existence of the global uniform attractor we present in Section \ref{section:single} a further step of the  proof of Theorem \ref{theorem:uniquesolution}. In detail,  we verify in Section \ref{section:fractal}, that this uniform attractor has a finite fractal dimension.
 We emphasize that this is in sharp contrast to the Hamiltonian conservative dynamics of the unperturbed DNLS for which the (stationary) breather solutions are supposed to lie on infinite-dimensional tori in phase space.
Finally, we present the proof that the attractor consists of a single trajectory.

\section{Preliminaries}

Here we collect some definitions, notations, terminology and hypotheses needed in the forthcoming. 
For a functional setting we consider solutions $\psi=(\psi_n)_{n\in {\mathbb{Z}}}\in C^1(\mathbb{R};l^2)$, where $l^2$ is the Hilbert space of square-summable sequences:
\begin{equation}
 l^2=\left\{ \psi=(\psi_n)_{n \in {\mathbb{Z}}}\,\in {\mathbb{C}}\,\,\,\vert \, \parallel \psi\parallel_{l^2}=\left(\sum_{n \in {\mathbb{Z}}}|\psi_n|^2\right)^{1/2}<\infty\right\}.\nonumber
\end{equation}

Associated with $F$, we introduce the operator $F:\,l^2\rightarrow l^2$, which for every $\psi \in l^2$ is determined by 
\begin{equation}
\left(F(\psi)\right)_n=
F(|\psi_n|^2)\psi_n,\,\,\,n\in \mathbb{Z}.\nonumber
\end{equation}
$F$ is Lipschitz continuous on bounded sets of $l^2$.
In fact, let 
\begin{equation}
 \mathcal{B}_R:=\left\{ \psi \in l^2\,|\,
 \parallel \psi\parallel_{l^2}\le R\right\}\nonumber
\end{equation}
be the closed ball centered at $0$ of radius $R$  in $l^2$ and consider  $\psi \in \mathcal{B}_R$. 
Using (\ref{eq:A2}) we have for the nonlinear operator $F:l^2\rightarrow l^2$: 
\begin{eqnarray}
 \parallel F(\psi) \parallel_{l^2}^2&=& \sum_n\,|F(|\psi_n|^2)\psi_n|^2.\nonumber\\
 &\le&a^2R^{2b}\sum_n\,|\psi_n|^2=a^2R^{2b}||\psi||_{l^2}^2.\nonumber
\end{eqnarray}
Hence, $F:\,l^2\rightarrow l^2$ is bounded on bounded sets of $l^2$. 
For $\psi,\phi \in \mathcal{B}_R$ we derive 
\begin{eqnarray}
 \parallel F(\psi)-F(\phi)\parallel_{l^2}^2&=&
 \sum_{n}\left|F(|\psi_n|^2)\psi_n-|F(|\phi_n|^2)\phi_n\right|^2\nonumber\\
 &\le& 2a^2R^{2b}||\psi-\phi||_{l^2}^2=L||\psi-\phi||_{l^2}^2,\label{eq:Lipschitz}
\end{eqnarray}
verifying that the map $F:\,l^2\rightarrow l^2$ is Lipschitz continuous on bounded sets of $l^2$ with Lipschitz constant 
$L(a,b,R)=\sqrt{2}aR^b$. (In general, for any bounded set $\mathcal{B} \in l^2$ it holds $\parallel F(\psi)-F(\phi)\parallel_{l^2}\le C(\mathcal{B})||\psi-\phi||_{l^2}$.)

In  this paper we consider periodic, quasiperiodic and almost periodic drivings $g_0$, respectively.

\begin{definition} 
\label{definition:perodic}
A function $f(t)=\{f_n(t)\}_{n\in \mathbb{Z}}$ is said to be periodic in $t\in \mathbb{R}$ uniformly with respect to $n \in \mathbb{Z}$ if
there exists a $T>0$ such that 
\begin{equation}
 f_n(t+T)=f_n(t),\,\,\forall t\in \mathbb{R},\,\,\forall n\in \mathbb{Z}.\nonumber
\end{equation}

\end{definition}

\begin{definition} 
\label{definition:qp}
 Let $\omega_1,...,\omega_k$ be rationally independent, i.e.
 \begin{equation}
  \sum_{j=1}^k\,a_j\,\omega_j\neq 0,\,\,\,\forall a_j\in \mathbb{Q}\setminus \{0\}.\nonumber
\end{equation}

A function $f(t)=\{f_n(t)\}$ is said to be quasiperiodic in $t$ uniformly with respect to $n\in \mathbb{Z}$ if there exist nonzero rationally independent 
$\omega_1,...,\omega_k\in \mathbb{R}^k$ and a function $F_n(x)\in C(\mathbb{Z} \times \mathbb{R}^k)$ such that 
\begin{equation}
 f_n(t)=F_n(\omega_1t,...,\omega_kt)=F_n(\omega t)\label{eq:qp}
\end{equation}
for all $t\in \mathbb{R}$ and $n\in \mathbb{Z}$, where $\omega =(\omega_1,...,\omega_k)$, and
\begin{equation}
 F_n(x+2\pi e_j)=F_n(x),\,\,\forall x\in \mathbb{R}^k,\,\,\forall n\in \mathbb{Z}.\label{eq:qp1} 
\end{equation}
$e_j$, $j=1,...k$ is the unit vector of $\mathbb{R}^k$ with the $i$th component one and all others zero.\end{definition}

\begin{definition}\label{definition:almost}
 Let $(X,d)$ be a separable and complete metric space and let $f$ be a continuous mapping from $\mathbb{R}$ to $X$. The function $f$ is called almost periodic uniformly with respect to $n\in \mathbb{Z}$ if for every $\epsilon>0$, there exists a constant $k(\epsilon)$ such that any interval of length $k(\epsilon)$ contains at least a number $\tau$ for which
 \begin{equation}
  \sup_{t\in \mathbb{R}} d(f_n(t+\tau),f_n(t))<\epsilon,\,\,\,\forall n\in \mathbb{Z}.\nonumber
 \end{equation}

\end{definition}

We make the assumption:

\noindent {\bf{H2:}} $g_0$ and  $g_0$ are assumed to be almost periodic in $t$ uniformly with respect to $n \in \mathbb{Z}$.

\vspace*{0.5cm}

As almost periodic functions (including periodic and quasiperiodic ones) are bounded and uniformly continuous  on $\mathbb{R}$, they are in  $C_b(\mathbb{R};l^2)$, i.e. the space of bounded functions on 
$\mathbb{R}$ with values in $l^2$ and with norm 
\begin{equation}
 ||f(t)||_{C_b(\mathbb{R};l^2)}=\sup_{t\in \mathbb{R}}||f(t)||_{l^2}.\nonumber
\end{equation}
We set $g=(g,g)$.
Define  the translation operator $T(h)$ as:
\begin{equation}
 T(h)g(t)=g(t+h),\,\,t,h\in \mathbb{R}.\nonumber
\end{equation}
A function $g(t)$ is said to be translation compact in $[C_b(\mathbb{R};l^2)]^2$, if the set of translations $\{T(h)g=g(\cdot +h)\,|\,h\in \mathbb{R}\}$, is precompact in $[C_b(\mathbb{R};l^2)]^2$.
For a given $g_0\in [C_b(\mathbb{R};l^2)]^2$ denote the closure of the set of all its translations $\{T(h)g_0=g_0(\cdot+h)\,|\,h\in \mathbb{R}\}$ in $[C_b(\mathbb{R};l^2)]^2$ by $\mathcal{H}(g_0)$ (the hull of the function $g_0$).  Then for any $g\in \mathcal{H}(g_0)$, $g$ is almost periodic  and 
$\mathcal{H}(g)=\mathcal{H}(g_0)$. Further, $\{T(h)g\}_{h\in \mathbb{R}}$ forms a continuous translation group on $ \mathcal{H}(g_0)$ with the invariance property $T(h)\mathcal{H}(g_0)=\mathcal{H}(g_0)$ for all $h\in \mathbb{R}$.

For any $\psi \in l^2$ we introduce the linear operators $A,B,B^{*}:\,l^2 \rightarrow l^2$,
\begin{equation}
 (A\psi)_{n }=\psi_{n+1}-2\psi_n+\psi_{n-1},\nonumber
\end{equation}
\begin{equation}
 (B\psi)_{n}=\psi_{n+1}-\psi_n,\qquad (B^{*}\psi)_{n}=\psi_{n-1}-\psi_n, \nonumber
\end{equation}
satisfying
\begin{equation}
 (B\psi,\theta)_{l^2}=(\psi,B^{*}\theta)_{l^2},\,\,\,\forall \psi,\theta\in l^2,\nonumber
\end{equation}
and  $-A=BB^{*}=B^{*}B$ from which one infers that 
\begin{equation}
 (A\psi,\psi)_{l^2}=-\parallel B\psi\parallel_{l^2}\le 0,\,\,\,\forall \psi \in l^2.\nonumber
\end{equation}
The operator $A$ is bounded according to
\begin{equation}
 \parallel A\psi\parallel_{l^2}\le 4\parallel \psi\parallel_{l^2}.\label{eq:boundDelta}
\end{equation}

Since $D(A)=D(A)^*=l^2$, and $A=A^*$, the operator $iA\,:\,l^2\mapsto l^2$ defined by $(iA)\psi=iA\psi$ for $\psi\in l^2$ is $\mathbb{C}$-linear and skew-adjoint and $iA$ generates a group $(\mathcal{T}(t))_{t\in \mathbb{R}}$ of isometries on $l^2$.

\section{Global existence and a family of processes} \label{section:existence}
In this section we  treat  the IVP for the infinite system of ordinary differential  (\ref{eq:system}),
(\ref{eq:icsystem})  
and prove the existence of global and unique solutions. Furthermore, we attribute a family of processes to the system (\ref{eq:system}),(\ref{eq:icdotpsi}) and verify the continuity of these processes with respect to initial data $\psi_{t_0}$
and $g\in \mathcal{H}(g_0)$.

The IVP (\ref{eq:system}),(\ref{eq:icsystem}) is equivalent to the following abstract nonautonomous LDS in the Hilbert space $l^2$:
\begin{equation}
  i\dot{\psi}-\kappa A\psi+i\gamma\psi +F(|\psi|^2)\psi-g=0,\,\,\,t>t_0,\label{eq:dotpsi}
\end{equation}
with given sequences $g(t)$ and $g(t)$, 
and initial data
\begin{equation}
 \psi(t_0)=\psi_{t_0}\in l^2.\label{eq:icdotpsi}
\end{equation}

 For  $\psi_{t_0}\in l^2$ a function $\psi\in C([t_0,T_0];l^2)$ is a solution to (\ref{eq:dotpsi}),(\ref{eq:icdotpsi}) for fixed $T_0>0$ iff  
\begin{eqnarray}
 \psi(t)&=&\mathcal{T_0}(t)\psi_{t_0}+i\int_{t_0}^T\,\mathcal{T}(t-\tau)f(\psi(\tau))d\tau,\nonumber\\
f(\psi)&=&i\gamma\psi +F(|\psi|^2)\psi-g.\nonumber
 \end{eqnarray}

Regarding the global existence of a unique solution to (\ref{eq:dotpsi}),(\ref{eq:icdotpsi})  we have the following:

\begin{proposition}\label{proposition:globalsolution}
Assume {\bf{H1}} and {\bf{H2}}. Let $g_0 \in [C_b(\mathbb{R},l^2)]^2$, $g\in \mathcal{H}(g_0)$, $t_0\in \mathbb{R}$ and $\psi_{t_0} \in l^2$ with $||\psi_{t_0}||_{l^2}\le r$.  
Then for every $\psi_{t_0}\in l^2$ 
the system (\ref{eq:system}) possesses a unique global solution $\psi(t)$ on $\mathbb{R}$ belonging to
$C^1(\mathbb{R};l^2)$.\label{Proposition:unique}
\end{proposition}

\noindent{\bf Proof:} 
Since the nonlinear operator $F$ is Lipschitz continuous on  bounded sets of $l^2$, and as due to (\ref{eq:boundDelta}) $A$ is a bounded linear operator on $l^2$, for any given initial data $\psi_{t_0}\in l^2$  
the local existence of a unique solution $\psi(t)\in C^1([t_0,T_0);l^2)$ for some $T_0>0$ can be verified by standard methods from the theory of ODEs (see \cite{Cazenave}-\cite{Zeidler1}).  
Whenever $T_0<\infty$ then 
$\lim_{t\rightarrow T_0^{-}}\parallel \psi(t)\parallel_{l^2}=\infty$. 

\vspace{0.5cm}
To ensure the global existence of the solutions, that is, $T_0=\infty$,
 we multiply (\ref{eq:system}) by $-i\psi_n$ and sum over $n$ yielding
\begin{eqnarray}
 \frac{d}{dt} \parallel\psi\parallel_{l^2}^2  &=&2\sum_{n\in{\mathbb{Z}}}\left(\left[{\rm Re} \psi_n {\rm Im} g_n-
 {\rm Re} g_n {\rm Im} \psi_n\right]
  -\gamma|\psi_n|^2\right)\nonumber\\
&\le& -2\gamma\sum_{n\in{\mathbb{Z}}} |\psi_n|^2+4\sum_{n\in{\mathbb{Z}}}|g_n||\psi_n|\nonumber\\
&\le& -\gamma ||\psi||_{l^2}^2+ \frac{1}{\gamma}\parallel g\parallel_{l^2}^2,\,\,\,\forall t\ge t_0,\nonumber
\end{eqnarray}
where we used Young's inequality.
We arrive at the inequality 
\begin{equation}
 \frac{d}{dt} \parallel \psi(t)\parallel_{l^2}^2+{\gamma} ||\psi(t)||_{l^2}^2\le \frac{1}{{\gamma}}\parallel g(t)\parallel_{l^2}^2,\label{eq:inequ}
\end{equation}
from which utilising Gronwall's inequality we obtain
\begin{eqnarray}
 \parallel\psi(t)\parallel_{l^2}^2&\le& \parallel\psi_{t_0}\parallel_{l^2}^2\exp(-{\gamma} (t-t_0))+\frac{1}{{\gamma}}\int_{t_0}^t\,\parallel g(\tau)\parallel_{l^2}^2\exp(-{\gamma}(t-t_0))d\tau\nonumber\\
 &\le& \parallel\psi_{t_0}\parallel_{l^2}^2\exp(-{\gamma}(t-t_0))+\frac{1}{{\gamma}^2} \sup_{t\ge t_0}\parallel g(t)\parallel_{l^2}^2(1-\exp(-{\gamma}(t-t_0))).\nonumber
\end{eqnarray}
As $g\in \mathcal{H}(g_0)$ it holds $||g||_{C_b(\mathbb{R},l^2)}=
||g_0||_{C_b(\mathbb{R},l^2)}$, and we get 
\begin{equation}
 \parallel\psi(t)\parallel_{l^2}^2\le  
 \parallel\psi_{t_0}\parallel_{l^2}^2
 \exp(-\gamma_0(t-t_0))
 +\frac{1}{\gamma_0^2} \parallel g_0\parallel_{C_b(\mathbb{R},l^2)}^2.\label{eq:boundedpsi}
\end{equation}
Crucially, the right hand side of the inequality (\ref{eq:boundedpsi}) holds for all $t\ge t_0\in \mathbb{R}$ and does not depend on $g\in \mathcal{H}(g_0)$. Hence,  $\parallel \psi(t)\parallel_{l^2}^2
< \infty$ for all $t\in \mathbb{R}$.

For $\parallel \psi\parallel_{l^2}^2
\le R$, that is on bounded sets in $l^2$,  we derive  from (\ref{eq:system}) 
\begin{equation}
 \frac{d}{dt}||{\psi}(t)||_{l^2}\le (4\kappa+aR^b+\gamma+ \parallel g_0\parallel_{C_b(\mathbb{R},l^2)}<\infty,\nonumber
\end{equation}
assuring that the solution belongs to $C^1(\mathbb{R};l^2)$.

\ \ \ $\square$

\vspace*{0.5cm}
In conclusion, we observe that for every $g\in \mathcal{H}(g_0)$, $t_0\in \mathbb{R}$ and $\psi_{t_0} \in l^2$ the solution $\psi$ to (\ref{eq:dotpsi}),(\ref{eq:icdotpsi}) exists so that a family of processes  $\{U^g(t,t_0)|\,t\ge t_0,\,t_0 \in \mathbb{R}\}_{g\in \mathcal{H}(g_0)}$ on $l^2$, with time symbol $g\in \mathcal{H}(g_0)$,  can be assigned to system (\ref{eq:dotpsi}),(\ref{eq:icdotpsi}).  In detail, if $\psi_{t_0}\in l^2$, then $U^g(t_0,t)\psi_{t_0}=\psi(t)$ determines the state of the system  (\ref{eq:dotpsi}),(\ref{eq:icdotpsi}) at time $t$  for given $g\in \mathcal{H}(g_0)$, $t_0\in \mathbb{R}$ and $t\ge t_0$. 
Moreover, as the IVP (\ref{eq:dotpsi}),(\ref{eq:icdotpsi}) has a unique solution, the family of processes $\{U^g(t,t_0)\}_{g\in \mathcal{H}(g_0)}$ possesses the multiplicative features:
\begin{eqnarray}
 U^g(t,s)U^g(s,t_0)&=&U^g(t,t_0),\,\,\,\forall t_0 \le s \le t,\,\,\,t_0 \in \mathbb{R},\nonumber\\
 U^g(t_0,t_0)&=&\mathbb{I},\nonumber
\end{eqnarray}
where $\mathbb{I}$ is the identity operator.

Next we show that the continuity of the processes with respect to initial data $\psi_{t_0}$
and $g\in \mathcal{H}(g_0)$.

\begin{lemma}\label{Lemma:continuity}
 Assume $g_0 \in [C_b(\mathbb{R},l^2)]^2$, {\bf H1}. and {\bf{H2}}.
 Let $g,\,g_n \in \mathcal{H}(g_0)$ and $\Theta,\,\Theta_n \in l^2$ for all $n\in \mathbb{Z}_+$, where $\mathbb{Z}_+=1,2,...$. Suppose $g_n \rightarrow g$ and $\Theta_n\rightarrow \Theta$ as $n\rightarrow \infty$. Then for every $t\ge t_0$ one has 
 \begin{equation}
  \lim_{n\rightarrow \infty}||U^{g_n}(t,t_0)\Theta_n-U^g(t,t_0)\Theta||_{l^2}=0.\nonumber
 \end{equation}
\end{lemma}

{\bf Proof:} Denote $\alpha_n(t,t_0)=U^{g_n}(t,t_0)\Theta_n$, $\alpha(t,t_0)=U^{g}(t,t_0)\Theta$
 and $\Delta_n(t,t_0)=\alpha_n(t,t_0)-\alpha(t,t_0)$.
From  (\ref{eq:dotpsi}) we derive 
\begin{equation}
i \frac{d \Delta_n}{dt}=\kappa A\Delta_n-i\gamma\Delta_n+F(|\alpha_n|^2)\alpha_n-F(|\alpha|^2)\alpha+g_n-g=0.\nonumber
\end{equation}
and by taking the inner product with $i\Delta_n$ we get
\begin{equation}
 \frac{1}{2}\frac{d}{dt} \parallel\Delta_n\parallel_{l^2}^2  =-i\kappa (A\Delta_n,\Delta_n)+\gamma||\Delta_n||_{l^2}^2
 +i(F(|\alpha_n|^2)\alpha_n-F(|\alpha|^2)\alpha,\Delta_n)
 -i(g_n-g,\Delta_n)=0.\nonumber
\end{equation}
Using (\ref{eq:Lipschitz}) we estimate
\begin{equation}
 |(F(|\alpha_n|^2)\alpha_n-F(|\alpha|^2)\alpha,\Delta_n)|\le ||F(|\alpha_n|^2)\alpha_n-F(|\alpha|^2)\alpha||_{l^2} ||\Delta_n||_{l^2}\le L ||\Delta_n||_{l^2}. \nonumber
\end{equation}

With the help of the properties of the linear operator $A$ and Young's inequality we get
\begin{eqnarray}
 \frac{d}{dt} \parallel\Delta_n\parallel_{l^2}&\le& (\gamma+L+4\kappa)\parallel\Delta_n\parallel_{l^2}+
 ||g_n-g||_{l^2},\nonumber
\end{eqnarray}
giving with Gronwall's inequality
\begin{eqnarray}
 \parallel\Delta_n(t,t_0)\parallel_{l^2}&\le& \exp[(\gamma+L+4\kappa)(t-t_0)]\parallel\Delta_n(t_0,t_0)\parallel_{l^2}^2+\int_{t_0}^t\exp[(\gamma+L+4\kappa)(t-t_0)]\nonumber\\
 &\times& \left(||g_n-g||_{l^2}\right) d\tau\nonumber\\
 &\le& \exp[(2(\gamma+L+4\kappa))(t-t_0)]\parallel\Theta_n(t_0)-\Theta(t_0)\parallel_{l^2}^2+\frac{\exp[2(\gamma+L+4\kappa)(t-t_0)]}{2(\gamma+L+4\kappa)}\\\nonumber
 &\times&\left(||g_n-g||_{C_b(\mathbb{R},l^2)}
 +||\alpha_n-\alpha||_{C_b(\mathbb{R},l^2)}\right)
 \rightarrow 0,\,\,\,{\rm as}\,\,n\rightarrow \infty\nonumber
\end{eqnarray}
completing the proof.

\ \ \ $\square$

\section{Breather solutions}\label{section:localised}
After having established the existence of a global unique solution to the IVP (\ref{eq:system}),(\ref{eq:icsystem}) we consider now    localised solutions in the form of  breather solutions whose modulus oscillates.  

\subsection{Periodic breather solutions}\label{subsection:periodic}
In this subsection the external driving field is supposed to be time-periodic in $t\in \mathbb{R}$, that is $g_n(t+T)=g_n(t)$ for $T>0$ uniformly with respect to all $n$. We discuss the existence of  breathers as spatially localised and time-periodic   solutions to the nonautonomous  system (\ref{eq:system}). Such solutions   are characterised by:
 \begin{eqnarray}
  \psi_n(t+T)&=&\psi_n(t),\,\, n\in {\mathbb{Z}},\label{eq:periodic}\\
  \lim_{|n|\rightarrow \infty}|\psi_n|&=&0,\label{eq:limits}
 \end{eqnarray}
 for some period $T>0$.

Regarding  the existence of time-periodic breather solutions we state:

\begin{proposition} 
Assume $g(t)$ is time-periodic with period $T$. Then there exists a unique time-periodic solution  periodic solution $\psi(t)\in C^1(\mathbb{R};l^2)$ of (\ref{eq:system}) of the same period $T$.
\end{proposition}

\noindent {\bf Proof:} By contradiction: Suppose that $\psi(t)\neq \psi(t+T)$.   By the proven existence of a unique global solution  in Proposition \ref{Proposition:unique},  one has that $\psi(t+T)$ is a solution with  driving term $g(t+T)$. Then the uniqueness of the solution and $\psi(t)\neq \psi(t+T)$ imply $g(t)\neq g(T+t)$ contradicting the periodicity  $g(t)=g(t+T)$.  Hence,
$\psi(t)=\psi(t+T)$ and the proof is finished.

\ \ \ $\square$

\vspace*{0.5cm}

We remark that the  decay of the states for $|n|\rightarrow \infty$ takes place in the sense of the 
$l^2$ norm. To prove the existence of exponentially localised (single-site) breathers  appropriately (e.g. exponentially) weighted function spaces can be used (see in \cite{JMP}).

\subsection{Quasiperiodic breathers}
\label{subsection:quasiperiodic}
Here we establish the existence of quasiperiodic breathers for quasiperiodic external driving field $g$.

\begin{proposition} Assume $g(t)$ is quasiperiodic in the sense of Definition \ref{definition:qp}. Then there exists a unique quasiperiodic solution $\psi(t)\in C^1(\mathbb{R};l_w^2)$ of system (\ref{eq:system}).
\end{proposition}

\noindent {\bf Proof:} The proof facilitates ideas originating from studies of continuous systems   in \cite{qp1},\cite{qp2}, which we readily  adapt to be applied to our discrete systems. For quasiperiodic $g(t)=(g_n(t))_{n\in \mathbb{Z}}$ there exists $G_n(x)$ satisfying  
 \begin{equation}
  G_n(x+2\pi e_j)=G_n(x),\,\,\forall x\in \mathbb{R}^k\,\,{\rm and}\,\,n\in \mathbb{Z},\nonumber
 \end{equation}
 and 
 \begin{equation}
  g_n(t)=G_n(\omega_1t,...,\omega_kt)=G_n(\omega t), \nonumber
 \end{equation}
 for all $t\in \mathbb{R}$ and $n\in \mathbb{Z}$.
We use the notations $\tilde{g}_n(t)=G_n(\omega t+\tilde{\alpha})$, with $\tilde{\alpha}=(\alpha_1,...,\alpha_k)$, and $\psi_{\tilde{\alpha}}(t)$ for the solution to (\ref{eq:system}) under the external forcing $\tilde{g}(t)$. 
Further, we introduce $\Gamma(\tilde{\alpha})=\psi_{\tilde{\alpha}}(0)$ and note, 
that   $\psi_{\tilde{\alpha}}(t+h)$ and  $\psi_{\tilde{\alpha}+\omega h}(t)$ is the unique solution to  (\ref{eq:system}) with external driving $\tilde{g}(t+h)$ and $G(\omega (t+h)+\tilde{\alpha})=\tilde{g}(t+h)$, respectively.
By the properties of the translation group, one has 
\begin{equation}
 \psi_{\tilde{\alpha}}(t+h)=\psi_{\tilde{\alpha}+\omega h}(t),\nonumber
\end{equation}
which for $t=0$ reads $\psi_{\tilde{\alpha}}(h)=\Gamma(\tilde{\alpha}+\omega h)$. Hence,  $\psi_{\tilde{\alpha}}(t)=\Gamma(\tilde{\alpha}+\omega t)$ for all $t\in \mathbb{R}$. Further, one has 
\begin{equation}
 \Gamma (\tilde{\alpha}+2\pi e_j)=\psi_{\tilde{\alpha}+2\pi e_j}(0),\,\,j=1,...k,\nonumber
\end{equation}
and $\psi_{\tilde{\alpha}+2\pi e_j}(t)$ is the unique solution of (\ref{eq:system}) with external driving $G(\omega t+\tilde{\alpha}+2\pi e_j)=G(\omega t +\tilde{\alpha})$,
and by the uniqueness of the solution we have 
$\psi_{\tilde{\alpha}+2\pi e_j}(t)=\psi_{\tilde{\alpha}}(t)$, which reads for $t=0$ as $\psi_{\tilde{\alpha}+2\pi e_j}(0)=\psi_{\tilde{\alpha}}(0)$. Hence,
\begin{equation}
 \Gamma(\tilde{\alpha}+2\pi e_j)=\Gamma(\tilde{\alpha}),\,\,j=1,...,k.\label{eq:qpf}
\end{equation}
Finally, for $\tilde{\alpha}=0$ in $\psi_{\tilde{\alpha}}(h)=\Gamma(\tilde{\alpha}+\omega h)$, we conclude 
that $\psi(t)=\Gamma(\omega t)$
and $G(\omega t +\tilde{\alpha})=G(\omega t)$.
 Then the relation (\ref{eq:qpf}) and the Definition \ref{definition:qp} imply  that $\psi(t)$ is quasiperiodic in $t$.

\ \ \ $\square$

We emphasize that the DNLS without damping and driving does not support quasiperiodic localised (i.e. square summable)  solutions on KAM tori. The reason is the presence of an absolutely  continuous spectrum (phonon band) lying within the interval $[0,4\kappa]$ in the corresponding linearised system.  
If a quasiperiodic solution with at least two incommensurable frequencies existed, there would be harmonics densely on the real axis generated having overlap with the interval of the absolutely continuous spectrum. This  causes radiation away from the center  of localisation towards infinity so that localised solutions cannot persist.  
However, with the presence of the linear damping term the absolutely continuous spectrum lies  no longer on the real axis so that  resonances between the phonon frequencies and  the frequencies of a quasiperiodic breather can be avoided.

\section{Dissipative dynamics--Absorbing sets and uniform attractors}\label{section:attractors}
\setcounter{equation}{0}
In this section we verify the existence of a  uniform  attractor for the nonautonomous LDS associated with the infinite system (\ref{eq:dotpsi}).(\ref{eq:icdotpsi}). We recall the definition of an absorbing set and a uniform attractor, respectively \cite{Babin},\cite{Chepyzhov1},\cite{Chepyzhov}.

 \begin{definition} 
 \label{def:absorbing}	
 A set $\mathcal{B}_0\in l^2$  is called uniformly (with respect to $g\in \mathcal{H}(g_0)$) absorbing for the family of processes $\{U^g(t,t_0)\}_{g\in \mathcal{H}(g_0)}$, if for every $t_0 \in \mathbb{R}$ and for all bounded sets $\mathcal{B} \subset l^2$ there exists $T(t_0,\mathcal{B})\ge t_0$ such that 
 \begin{equation}
 \bigcup \, U^g(t,t_0)\mathcal{B} \subset \mathcal{B}_0,\,\,\,\forall t\ge T(t_0,\mathcal{B}).
 \end{equation}
  \end{definition}

\begin{definition}
\label{definition:attractor}
A closed set ${\cal A}$ is called a uniform (with respect to $g\in \mathcal{H}(g_0)$) attractor  for the 
family of processes  $\{U^g(t,t_0)\}_{g\in \mathcal{H}(g_0)}$, if

  (i)  $\lim_{t\rightarrow \infty}\,\sup_{g\in \mathcal{H}(g_0)} dist(U^g(t,t_0)B,{\cal A})=0$, for each fixed $t_0 \in \mathbb{R}$ and each bounded set $\mathcal{B}\in l^2$, where the Hausdorff semi-distance between two nonempty subsets $U,V$ of $l^2$ is determined by
 \begin{equation}
  dist(U,V)=\sup_{u\in U}\,\inf_{v\in V}\,d(u,v)_{l^2};\nonumber
 \end{equation}
 (ii)  ${\cal A}$ is the minimal set among those satisfying (i), that is, if $\tilde{\mathcal{A}}$ is any closed set of $l^2$ having the property (i), then $\mathcal{A}\subseteq \tilde{\mathcal{A}}$.
\end{definition}

 First we establish the existence of a uniform absorbing set in $l^2$ for the family of processes  $\{U^g(t,t_0)\}_{g\in \mathcal{H}(g_0)}$.
In particular, we derive bounds for the solutions of the LDS that are uniform with respect to $g\in \mathcal{H}(g_0)$.

\begin{lemma}
\label{Lemma:absorbingmu}	
Assume $g_0\in C_b(\mathbb{R},l^2)$, {\bf H1} and {\bf{H2}}.  For the family of processes $\{U^g(t,t_0)\}_{g\in \mathcal{H}(g_0)}$ there exists a bounded absorbing set $\mathcal{B}_0\in l^2$, that is  for every $t_0 \in \mathbb{R}$ and for all bounded sets $\mathcal{B} \subset l^2$ there exists $T(t_0,\mathcal{B})\ge t_0$ such that 
 \begin{equation}
 \bigcup \, U^g(t,t_0)\mathcal{B} \subset \mathcal{B}_0,\,\,\,\forall t\ge T(t_0,\mathcal{B}).
 \end{equation}

Then there exists a constant $K$ depending only on $\gamma$ and $g_0$, such that  for any solution $\psi$ of the IVP (\ref{eq:dotpsi}),(\ref{eq:icdotpsi}) with $||\psi_{t_0}||_{l^2}\le r$, one has
\begin{equation}
 ||\psi(t)||_{l^2}\le K,\,\,\,\forall t\ge t_0+T,\label{eq:K}
\end{equation}
where $T$ depends only on $\gamma$, $g_0$ and $r$.
That is, the ball $\mathcal{B}_{0}$ of $l^2$ centered at $0$ of radius $K$, is a uniform absorbing set for the family of processes $\{U^g(t,t_0)\}_{g\in \mathcal{H}(g_0)}$. 
\end{lemma}

\noindent{\bf Proof:} 
With $||\psi_{t_0}||_{l^2}\le r$, the relation  (\ref{eq:boundedpsi})  implies  that
\begin{equation}
 || \psi(t)||_{l^2}^2\le 
 \exp(-\gamma(t-t_0))r^2
 +\frac{1}{\gamma^2} \parallel g_0\parallel_{C_b(\mathbb{R},l^2)}^2\le \frac{2}{\gamma^2} \parallel g_0\parallel_{C_b(\mathbb{R},l^2)}^2,\,\,\,{\rm for}\,\,t-t_0\ge T=\frac{1}{\gamma}\ln\,\frac{\gamma^2r^2}{\parallel g_0\parallel_{C_b(\mathbb{R},l^2)}^2}.\label{eq:absset}
\end{equation}
Hence there exists a constant $K$ depending only on $\gamma$ and $g_0$, such that  for any solution $\psi$ of the IVP (\ref{eq:dotpsi}),(\ref{eq:icdotpsi}) with $||\psi_{t_0}||_{l^2}\le r$, holds
\begin{equation}
 ||\psi(t)||_{l^2}\le K,\,\,\,\forall t\ge t_0+T,\nonumber
\end{equation}
where $T$ depends only on $\gamma$, $g_0$ and $r$, implying that  for any bounded set  
 $\mathcal{B}\in l^2$,   it follows that
 \begin{equation}
  U^g(t,t_0)\mathcal{B}\subseteq \mathcal{B}_0:=\{\psi \in l^2\,|\,||\psi||_{l^2}\le K\},\,\,\,\forall g\in \mathcal{H}(g_0)\,\,{\rm and}\,\,t-t_0\ge T,
 \end{equation}
and $T$ depends only on $\gamma$, $\mathcal{B}$ and $g_0$. Therefore, $\mathcal{B}_{0}$ is a uniform absorbing set for the family of processes $\{U^g(t,t_0)\}_{g\in \mathcal{H}(g_0)}$ 
concluding the  proof.

\ \ \ $\square$

\vspace*{0.5cm}

\begin{remark}
 Note that the absorbing set attracts any bounded region of $l^2$ exponentially fast.
\end{remark}

Next, we verify that solutions of the LDS   (\ref{eq:dotpsi}),(\ref{eq:icdotpsi}) possess the asymptotic tail-end property 
 (see \cite{Bates,Chepyzhov1,Chepyzhov,Hale, Zhou0,Zhou,Zhou1,Zhou3,Wang,Zhou2,Nikos,NAE2007,Du}). We stress that the asymptotic tail property is essential in certifying the asymptotic compactness of the processes.

\begin{lemma}
\label{Lemma:asymtailmu}	
 Assume $g_0\in C_b(\mathbb{R},l^2)$, {\bf H1} and {\bf{H2}}. Let $||\psi_{t_0}||_{l^2} \le r$. Then, for any $\xi>0$, there exist
 $T(\xi,r)$ and $M(\xi)\in \mathbb{N}$ such that the solution $\psi(t)$ of (\ref{eq:dotpsi})(\ref{eq:icdotpsi}) satisfies for all $t-t_0\ge T(\xi)$:
 \begin{equation}
  \sum_{|n|> K}|\psi_n(t)|^2\le \xi\,\,\,\,{\rm for\, any\,\,\,\,} K\ge M(\xi).\label{eq:asymptotic}
 \end{equation}
\end{lemma}

\noindent{\bf Proof:}
Since the solution of (\ref{eq:dotpsi}) satisfies the relation (\ref{eq:inequ}), the latter is satisfied by the solution $\psi_k(t)$ for all $k\in\mathbb{Z}$, in particular for all $|k|>m$ for some $m\in\mathbb{N}$. 
Similarly as in the proof of Proposition  \ref{Proposition:unique} we derive the differential inequality 
\begin{eqnarray}
  \frac{d}{dt}\sum_{|k|>m}|\psi_k(t)|^2\le 
   -\gamma\sum_{|k|>m}|\psi_k(t)|^2+\frac{1}{\gamma}\sum_{|k|>m}|g_k(t)|^2,\,\,\,{\rm for}\,\, t>t_0,
  \end{eqnarray}
  which by Gronwall's inequality yields
\begin{eqnarray}
 \sum_{|k|>m}|\psi_{k}(t)|^2 &\le& \sum_{|k|>m}|\psi_{k}(t_0)|^2 \exp(-\gamma(t-t_0))
+\frac{1}{\gamma^2}\int_{t_0}^t
\sum_{|k|>m} |g_{k}(\tau)|^2\left(1-\exp(-\gamma(\tau-t_0))\right)d\tau\nonumber\\
&\le&\sum_{|k|>m}|\psi_{k}(t_0)|^2 \exp(-\gamma(t-t_0))
+\frac{1}{\gamma^2}\sup_{t\ge t_0}
\sum_{|k|>m} |g_{k}(t)|^2.\label{eq:partialsum1}
 \end{eqnarray}
 We treat the two terms on the r.h.s. of (\ref{eq:partialsum1}) separately.
Note that $\sum_{|k|>m}|\psi_{k}(t_0)|^2\le ||\psi_{t_0}||_{l^2}^2\le r^2$. Then for the first term we find that for any $\xi>0$ there is a $T(\xi,r)=\ln(2r^2/\xi)/\gamma$ such that 
\begin{equation}
 \sum_{|k|>m}|\psi_{k}(t_0)|^2 \exp(-\gamma(t-t_0))\le r^2 \exp(-\gamma(t-t_0))\le  \frac{\xi}{2}, \qquad\forall t-t_0\ge T(\xi,r).\label{eq:Txi}
\end{equation}
From $g_0$ being almost periodic follows that any $g\in \mathcal{H}(g_0)$ is almost periodic too. Thus, for a fixed $g\in \mathcal{H}(g_0)$ the set $\{g(t)=(g_n(t))_{n\in \mathbb{Z}}\,|\,t\in \mathbb{R}\}$ is precompact in $l^2$. Therefore for any $\xi>0$ there exists a constant  $M_1(g,\xi)$ such that 
\begin{equation}
 \frac{1}{\gamma^2}
\sum_{|k|>m} |g_{k}(t)|^2 \le \frac{\xi}{2}, \qquad{\rm for\,\,\,any}\,\, m\ge M_1(g,\xi)\,\,\,{\rm and}\,\,\,\forall t\ge t_0.
\end{equation}
Since $\mathcal{H}(g_0)$ is compact in $C_b(\mathbb{R},l^2)$, one finds a $M(\xi)$, that does not depend on $g$, such that for all $m\ge M(\xi)$
\begin{equation}
 \sum_{|k|>m}|g_{k}(t)|^2\le \frac{\xi}{2},\qquad \forall t\in \mathbb{R}\,\,\,{\rm and\,\, all}\,\,g\in \mathcal{H}(g_0).\label{eq:Kxi}
\end{equation}

Combining (\ref{eq:Txi}) and (\ref{eq:Kxi}) gives
\begin{equation}
\sum_{|k|>m}|\psi_{k}(t)|^2 \le \xi,\qquad {\rm for\,\, all}\,\,t-t_0\ge T(\xi,r)\,\,{\rm and}\,\,m\ge M(\xi),
\end{equation}
and the proof is finished.

\ \ \ $\square$

\vspace*{0.5cm}

In order to establish the existence of a uniform attractor for the system (\ref{eq:dotpsi}),(\ref{eq:icdotpsi}) we introduce in the next step a semigroup of nonlinear operators for the family of processes $\{U^g(t,t_0)\}_{g\in \mathcal{H}(g_0)}$ in an extended phase space (see \cite{Chepyzhov1},\cite{Chepyzhov}). Subsequently, using the semigroup theory, it is shown that the semigroup is asymptotically compact and point dissipative so that it possesses a global attractor from  which one infers on the existence of a uniform attractor for the processes.

In the extended phase space $l^2 \times \mathcal{H}(g_0)$  we introduce a nonlinear semigroup $\{S(t)\}_{t\ge 0}$ as
\begin{equation}
 S(t)(\psi,g)=(U^g(t,0)\psi,T(t)g),\,\,{\rm for \,\,every}\,\,t\ge 0,\,\,\psi\in l^2,\,\,g\in \mathcal{H}(g_0),
\end{equation}
with $T$ as the translation group determined by $T(h)g=g(\cdot +h)$ for every $g\in \mathcal{H}(g_0)$.
Evidently, $\{T(h)\}_{h\in \mathbb{R}}$ forms a continuous translation group on $\mathcal{H}(g_0)$ leaving the latter invariant, that is
\begin{equation}
 T(h)\mathcal{H}(g_0)=\mathcal{H}(g_0),\,\,{\rm for\,\,all}\,\, h \in \mathbb{R},
\end{equation}

Crucially, $\{S(t)\}_{t\ge 0}$ is characterised by
\begin{equation}
 S(t)S(u)=S(t+u),\,\,\,S(0)=\mathbb{I},
\end{equation}
that is, it has the semigroup features.
From \cite{Chepyzhov1},\cite{Chepyzhov} it is known that if $\{S(t)\}_{t\ge 0}$ possesses a global attractor in the extended phase space $l^2 \times \mathcal{H}(g_0)$, then the family of processes $\{U^g(t,t_0)\}_{g\in \mathcal{H}(g_0)}$ possesses a uniform attractor in the phase space $l^2$ as a result of the projection of the global attractor of $\{S(t)\}_{t\ge 0}$ onto $l^2$.

Furthermore, one needs the notion of asymptotic compactness of a  semigroup \cite{Chepyzhov1}:

\begin{definition} 
\label{def:asymcompact}	
The semigroup  $\left\{S(t)\right\}_{t\ge 0}$ is said to be asymptotically compact in  $l^2\times \mathcal{H}(g_0)$ if  for any bounded $\mathcal{D} \subset l^2\times \mathcal{H}(g_0) $, and any sequence 
$\left\{t_n\right\}_{n\in \mathbb{Z}_+}$,  $\left\{\psi_n,g_n\right\}_{n\in \mathbb{Z}_+}$ with $t_n\ge 0$, $t_n \rightarrow \infty$ as $n\rightarrow \infty$, and $\psi_n,g_n \in \mathcal{D}$, the sequence  $\left\{S(t_n)(\psi_n,g_n)\right\}_{n\in \mathbb{Z}_+} $ is relatively compact in $l^2\times \mathcal{H}(g_0)$.
\end{definition}

The detailed description of the structure of the uniform attractor is based on the notion of a complete solution defined as follows \cite{Chepyzhov1}:

\begin{definition} 
\label{def:complete}
A curve $t\rightarrow \psi(t)\in l^2$ is called a complete solution for the process $U^g(t,t_0)$  for given $g\in \mathcal{H}(g_0)$, if
\begin{equation}
 U^g(t_0,t)\psi(t_0)=\psi(t),\,\,\,\forall t_0 \in \mathbb{R}\,\,{\rm and}\,\,t\ge t_0.\label{eq:kernel}
\end{equation}
The kernel of the process $U^g(t,t_0)$ is the collection $\mathcal{K}_g$ of all its bounded complete solutions $\mathcal{K}_g=\{\psi(\cdot)\in C_b(\mathbb{R};l^2)\,|\,\psi(\cdot )\,\,{\rm fulfilles}\,\, (\ref{eq:kernel})\}$.

The kernel section of the process $U^g(t,t_0)$ at time $\tau\in \mathbb{R}$ is the set  $\mathcal{K}_g(\tau)=\{\psi(\tau)\,|\,\psi(\cdot ) \in \mathcal{K}_g$.
\end{definition}

Establishing the  existence of a global attractor of the semigroup $S(t)$  is based on the following statement \cite{Chepyzhov1},\cite{Chepyzhov}:

\begin{proposition}
\label{Proposition:SA}
Provided the semigroup $S(t)$ is continuous, point dissipative and asymptotically compact, it possesses a compact global attractor $\mathcal{A}_S \in l^2\times \mathcal{H}(g_0)$. Denote by $\mathcal{P}_{l^2}$ and  
$\mathcal{P}_{\mathcal{H}(g_0)}$ the projectors from   $l^2\times \mathcal{H}(g_0)$ onto 
$l^2$ and $\mathcal{H}(g_0)$, respectively. Then $\mathcal{A}=\mathcal{P}_{l^2} \mathcal{A}_S$ is a compact uniform attractor for the family of processes $\{U^g(t,t_0)\}_{g\in \mathcal{H}(g_0)}$, and
\\
\noindent (i) $\mathcal{A}_S =\bigcup_{g\in \mathcal{H}(g_0)} \mathcal{K}_g(0) \times \{g\}$,\\
(ii) $\mathcal{A} =\bigcup_{g\in \mathcal{H}(g_0)} \mathcal{K}_g(0)$,\\
(iii) $\mathcal{P}_{\mathcal{H}(g_0)} \mathcal{A}_S=\mathcal{H}(g_0)$.

\end{proposition}

Next we establish the asymptotic compactness of the semigroup $\{S(t)\}_{t\ge 0}$.

\begin{proposition}
Under the same conditions of  Lemma \ref{Lemma:asymtailmu} the semigroup $\left\{S(t)\right\}_{t\ge 0}$  is asymptotically compact.
\label{Proposition:asymcompmu}
\end{proposition}

\noindent{\bf Proof:}
Since the sequence $\{\psi_n(t_n)\}_{n\in \mathbb{Z}_+}$ is uniformly bounded in $l^2$, in conjunction with relation (\ref{eq:K}),  it follows that the set $\{U^{g_n}(t_n,0)\psi_n\}_{n\in \mathbb{Z}_+}$ is bounded in $l^2$ so that the latter is weakly compact. That is  there is a subsequence  $\{U^{g_n}(t_n,0)\psi_n\}_{n\in \mathbb{Z}_+}$ (not relabelled) and $\psi\in l^2$ such that 
$U^{g_n}(t_n,0)\psi_n(t_n)\rightarrow \psi$ weakly in $l^2$ as $n \rightarrow \infty$.
To prove strong convergence we note that for any $\epsilon>0$ by Lemma \ref{Lemma:asymtailmu} there  exists $M_1(\epsilon)$ and $K_1(\epsilon)$ such that
\begin{equation}
 \sum_{|k|>K_1(\epsilon)}|(U^{g_n}(t_n,0)\psi_n)_k|^2< \frac{\epsilon}{5}\qquad\forall n>M_1(\epsilon).
\end{equation}
Since $\psi\in l^2$, there exists $K_2(\epsilon)$ such that $\sum_{|k|>K_2(\epsilon)}|\psi_k|^2< \epsilon/5$. Let $K(\epsilon)=\max\{K_1(\epsilon),K_2(\epsilon)\}$. Due to the weak convergence we have $(U^{g_n}(t_n,0)\psi_n(t_n))_k\rightarrow \psi_k$ as $n \rightarrow \infty$ for $|k|\le K(\epsilon)$.
Hence, there exists $M_2(\epsilon)$ such that 
\begin{equation}
 \sum_{|k|\le K(\epsilon)}|(\psi_n(t_n))_k-\psi_k|^2< \frac{\epsilon}{5}\qquad\forall n\ge M_2(\epsilon).
\end{equation}
With $M(\epsilon)=\max\{M_1(\epsilon),M_2(\epsilon)\}$
and $n\ge M(\epsilon)$ we derive
\begin{eqnarray}
 ||U^{g_n}(t_n,0)\psi_n-\psi||_{l^2}^2&=&\sum_{|k|>K(\epsilon)}|(U^{g_n}(t_n,0)\psi_n)_k-\psi_k|^2+\sum_{|k|\le K(\epsilon)}|(U^{g_n}(t_n,0)\psi_n)_k-\psi_k|^2\nonumber\\
 &\le& 2\sum_{|k|>K(\epsilon)}|(U^{g_n}(t_n,0)\psi_n)_k|^2+2\sum_{|k|>K(\epsilon)}|\psi_k|^2+\frac{\epsilon}{5} \le \epsilon.\label{eq:Uepsilon}
\end{eqnarray}
Conclusively, $U^{g_n}(t_n,0)\psi_n \rightarrow \psi$ strongly in $l^2$ as $n \rightarrow \infty$.

Furthermore, due to the compactness of $\mathcal{H}(g_0)$ and $T(t_n)g_n \in \mathcal{H}(g_0)$, there exists a $\hat{g}\in \mathcal{H}(g_0)$ such that 
\begin{equation}
 T(t_n)g_n \rightarrow \hat{g}\,\,{\rm in }\,\,\mathcal{H}(g_0)\,\,{\rm as}\,\,n\rightarrow \infty,
\end{equation}
which together with (\ref{eq:Uepsilon}) yields
\begin{equation}
 S(t_n)(\psi_n,g_n)=(U^{g_n}(t_n)\psi_n,T(t_n)g_n)\rightarrow (\psi,\hat{g})\,\,{\rm as}\,\,n\rightarrow \infty.
\end{equation}

\ \ \ $\square$

\vspace*{0.5cm}

We are now able to state the main result of this section.

\begin{theorem}
Assume $g_0 \in C_b(\mathbb{R},l^2)$ is almost periodic. Then the family of processes  $\{U^g(t,t_0)\}_{g\in \mathcal{H}(g_0)}$ possesses a compact uniform (with respect to $g\in \mathcal{H}(g_0)$) attractor ${\cal A} \subset l^2$ in the sense of Definition \ref{definition:attractor}.\label{Theorem:attractorALS}
 \end{theorem}

\noindent{\bf Proof:}
The continuity of the processes with respect to the initial data and $g\in \mathcal{H}(g_0)$ established in Lemma \ref{Lemma:continuity}, in conjunction with the continuity of the translation group $\{T(t)\}_{t\in \mathbb{R}}$, imply that the semigroup $\left\{S(t)\right\}_{t\ge 0}$ is continuous in $l^2 \times \mathcal{H}(g_0)$.
The set $\mathcal{B}_S=\mathcal{B}_0\times \mathcal{H}(g_0)$, where $\mathcal{B}_0$ is the absorbing set of the processes $\{U^g(t,t_0)\}_{g\in \mathcal{H}(g_0)}$ defined in \ref{Lemma:absorbingmu}, is a bounded absorbing set for $\{S(t)\}_{t\ge 0}$. Thus,  $\{S(t)\}_{t\ge 0}$ is point attractive. Moreover, by Lemma \ref{Proposition:asymcompmu} $\{S(t)\}_{t\ge 0}$ is asymptotically compact so that with the aid of  Proposition \ref{Proposition:SA} we conclude that the family of processes possesses a uniform (with respect to $g\in \mathcal{H}(g_0)$) attractor $\mathcal{A}$ which is the union of all bounded trajectories of the processes, i.e. $\mathcal{A} =\bigcup_{g\in \mathcal{H}(g_0)} \mathcal{K}_g(0)$. The proof is  finished.
 
\ \ \ $\square$

\vspace*{0.5cm} 
 
\section{Strong dissipation}\label{section:strong}
We study the effect of strong damping on the dynamics of the damped and driven system (\ref{eq:system}). 
In particular  we aim to prove that for strong dissipation the global uniform attractor has a finite fractal dimension.  Moreover, for strong damping we shall illustrate that the global uniform attractor consists of a single trajectory of the underlying system. 
  \subsection{Finite fractal dimension of the uniform attractor}\label{section:fractal}
 In this section we show that the uniform attractor $\mathcal{A}$ for the family of processes 
 $\{U^g(t,t_0)\}_{g\in \mathcal{H}(g_0)}$
has a finite fractal dimension in $l^2$. We recall the definition of the fractal dimension of a compact set:
 \begin{definition} \cite{Chepyzhov1}
  Let $D$ be a compact set in a metric space $X$. The fractal dimension $\dim_f(D)$ is defined as
  \begin{equation}
   \dim_f(D)=\lim\,\sup_{\epsilon \rightarrow 0}\frac{\log N(D,\epsilon)}{-\log(\epsilon)},
  \end{equation}
  where $N(D,\epsilon)$ is the minimal number of closed balls of  radius $\epsilon$ required to cover $D$.
 \end{definition}
 
 \begin{proposition} \cite{Chepyzhov1}
 \label{proposition:finite}
  Let $X$ be a separable Hilbert space and $D$  a bounded closed set in $X$. Assume that there exists a mapping $G:\,D\mapsto X$ such that $D\subseteq GD$ and \\
  (i) G is Lipschitz continuous on $D$, i.e. there exists positive $M_0$ such that
  \begin{equation}
   ||G\phi_1-G\phi_2||_{X}\le M_0 ||\phi_1-\phi_2||_X,\qquad \phi_1,\phi_2\in D;
  \end{equation}
 (ii) there exist compact seminorms $n_1(u)$ and $n_2(u)$ on $X$ such that 
 \begin{equation}
  ||G\phi_1-G\phi_2||_{X}\le \nu ||\phi_1-\phi_2||_X+L[n_1(\phi_1-\phi_2)+n_2(G\phi_1-G\phi_2)],\qquad \phi_1,\phi_2\in D,
 \end{equation}
 with $0<\nu<1$ and $L>0$.
 Then $D$ is a compact set in $X$ with a finite fractal dimension.
 \end{proposition}
 
 A seminorm $n(x)$ on $X$ is said to be compact iff 
 for any 
 bounded set $B\subset X$ there exists a sequence $(x_n)\subset B$ such that $n(x_m-x_n)\rightarrow 0$ as $m,n \rightarrow \infty$.
 
 \vspace*{0.5cm}
 
 We recall that when the initial data lie on  the attractor, $\phi(t_0) \in \mathcal{A}$, then for the associated solution $\phi(t)=(\phi_n(t))_{n\in \mathbb{Z}}$   it holds
 \begin{equation}
  ||\phi(t)||_{l^2}\le \frac{||g||_{C_b(\mathbb{R},l^2)}}{\gamma}\equiv C_{\mathcal{A}}, \qquad \forall t\ge t_0.\label{eq:rhofractal}
 \end{equation}
 
 We now present  the statement for the fractal dimension of the uniform attractor $\mathcal{A}$:
 
  \begin{theorem}
   \label{theorem:fractal}
   Let 
\begin{equation}
 \gamma>\left(a\left(||g||_{C_b(\mathbb{R},l^2)}\right)^b\right)^{1/(1+b)}. \label{eq:assfractal}
\end{equation}
   Then 
   the uniform attractor 
  $\mathcal{A}\subset l^2$ has finite fractal dimension.
   \end{theorem}

{\bf Proof:} For any $g\in \mathcal{H}(g_0)$ let $\phi_{t_0},\psi_{t_0}\in \mathcal{A}$ and $\phi_g(t)=U^g(t,t_0)\phi_{t_0}$,  $\psi_g(t)=U^g(t,t_0)\psi_{t_0}$ and denote by  $w_g(t)=\phi_g(t)-\psi_g(t)$ the difference of  two solutions. As the following analysis is independent of $g\in \mathcal{H}(g_0)$, in what follows the subscript $g$ will be  omitted.
 First we show that condition (i) of Proposition \ref{proposition:finite} is satisfied.
 Taking the inner product we have 
 \begin{eqnarray}
  \frac{d}{dt}||w(t)||_{l^2}^2&=& -\gamma\sum_{n}|\phi_n-\psi_n|^2 +i\sum_{n}(F(|\phi_n|^2)\phi_n-F(|\psi_n|^2))\psi_n)\overline{w}_n
  +c.c.\nonumber\\
  &\le& -2(\gamma-aC_{\mathcal{A}}^b) ||w(t)||_{l^2}^2,\qquad t\ge t_0,\nonumber
 \end{eqnarray}
 giving 
 \begin{equation}
  ||w(t)||_{l^2}^2\le \exp[-2(\gamma-aC_{\mathcal{A}}^b)\,(t-t_0)]||w(t_0)||_{l^2}^2,\nonumber
 \end{equation}
 that is
 \begin{equation}
  ||\phi(t)-\psi(t)||_{l^2}\le M_0 ||\phi(t_0)-\psi(t_0)||_{l^2},\qquad t\ge t_0,\label{eq:bounded1}
 \end{equation}
 where $M_0=M_0(a,b,c,t,t_0)$.
 
 \vspace*{0.5cm}
 We set $\tilde{\gamma}=\gamma -aC_{\mathcal{A}}^b$. By hypothesis (\ref{eq:assfractal}) one has $\tilde{\gamma}>0$. 
 As far as  condition (ii) of Proposition \ref{proposition:finite} is concerned one gets from
 \begin{equation}
  \frac{d}{dt}||w(t)||_{l^2}^2
  \le -4\tilde{\gamma}||w(t)||_{l^2}^2+2\tilde{\gamma}\sum_n |w_n(t)|^2, \qquad t\ge t_0,\nonumber
 \end{equation}
 that 
 \begin{eqnarray}
  ||w(t)||_{l^2}^2&\le& ||w(t_0)||_{l^2}^2\exp[-2\tilde{\gamma}\,(t-t_0)]+\int_{t_0}^t \tilde{\gamma}\sum_{n} |w_n(\tau)|^2\exp[-2\tilde{\gamma}(t-\tau)]d\tau\nonumber\\
  &\le& ||w(t_0)||_{l^2}^2\exp[-2\tilde{\gamma}\,(t-t_0)]+ 2\sum_{n} \max_{\tau\in [t_0,t]}|w_n^2(\tau)|^2,\nonumber
 \end{eqnarray}
 yielding for $T>t_0$
 \begin{equation}
  ||\phi(T)-\psi(T)||_{l^2}^2\le ||\phi(t_0)-\psi(t_0)||_{l^2}^2
   \exp[-2\tilde{\gamma}\,T]+2\sum_{n} \max_{t\in [t_0,T]}|\phi_n(t)-\psi_n(t)|^2.\nonumber
 \end{equation}
 We need to show that  $n(\phi)=\left(\sum_{n} \max_{t\in [t_0,T]}|\phi_n(t)|^2\right)^{1/2}$ is a compact seminorm on $l^2$.  
 For  a sequence $(\phi^n)_{n\in \mathbb{N}}\subset \mathcal{B}_{c}$ (the closed ball of radius $c$ centered at zero in $l^2$) we have 
 \begin{equation}
  n(\phi^m-\phi^n)=\left(\sum_{k} \max_{t\in [t_0,T]}|\phi_k^m(t)-\phi_k^n(t)|^2\right)^{1/2}.\label{eq:sumfractal}
 \end{equation}
 For each $n\in \mathbb{N}$,  $\phi^n\in \mathcal{B}_c \subset l^2$  so that there exists $N_1(\epsilon)\in \mathbb{N}$ such that 
 $\sum_{|k|>N_1(\epsilon)}|\phi_k^n(t)|^2<\epsilon^2/8$ for  $t\in[t_0,T]$ and  any $n\in \mathbb{N}$.
 We split the sum in (\ref{eq:sumfractal}) as
 \begin{eqnarray}
  \sum_{k} \max_{t\in [t_0,T]}|\phi_k^m(t)-\phi_k^n(t)|^2&=&\sum_{|k|\le N_1(\epsilon)} \max_{t\in [t_0,T]}|\phi_k^m(t)-\phi_k^n(t)|^2 +\sum_{|k|>N_1(\epsilon)} \max_{t\in [t_0,T]}|\phi_k^m(t)-\phi_k^n(t)|^2\nonumber\\
  &\le& \sum_{|k|\le N_1(\epsilon)} \max_{t\in [t_0,T]}|\phi_k^m(t)-\phi_k^n(t)|^2+\sum_{|k|> N_1(\epsilon)} \max_{t\in [t_0,T]}|\phi_k^m(t)|^2\nonumber\\
 &+&\sum_{|k|> N_1(\epsilon)} \max_{t\in [t_0,T]}|\phi_k^n(t)|^2+2\left(\sum_{|k|> N_1(\epsilon)} \max_{t\in [t_0,T]}|\phi_k^m(t)|^2\right)^{1/2}\,\left(\sum_{|k|> N_1(\epsilon)} \max_{t\in [t_0,T]}|\phi_k^n(t)|^2\right)^{1/2}
  \nonumber\\
  &\le&\sum_{|k|\le N_1(\epsilon)} \max_{t\in [t_0,T]}|\phi_k^m(t)-\phi_k^n(t)|^2+\frac{\epsilon^2}{2}.\label{eq:sumfractal1}
 \end{eqnarray}
 For the treatment of the finite sum in (\ref{eq:sumfractal1}), we note that by (\ref{eq:rhofractal}) 
 the solutions $\phi(t)$  with $\phi(t_0)\in \mathcal{A}$ are uniformly bounded and are uniformly continuous  on the closed interval $[t_0,T]$. Furthermore, $(\phi^n)_{n\in \mathbb{N}}\subset \mathcal{B}_c$ is weakly relatively compact.   Therefore there exists a $\phi\in \mathcal{B}_c$ such that $\phi^n\rightarrow \phi$ weakly in $B_c$ as $n\rightarrow \infty$. Since weak and strong convergence coincide in finite dimensional spaces one has that $\phi^n_k\rightarrow \phi_k$ strongly in $\mathbb{R}^{2N_1(\epsilon)+1}$. Hence,  there is $N_2(\epsilon)\in \mathbb{N}$ such that 
 \begin{eqnarray}
  \sum_{|k|\le N_1(\epsilon)} \max_{t\in [t_0,T]}|\phi_k^m(t)-\phi_k^n(t)|^2&=& \sum_{|k|\le N_1(\epsilon)} \max_{t\in [t_0,T]}|\phi_k^m(t)-\phi_k+\phi_k-\phi_k^n(t)|^2\nonumber\\
  &\le & \sum_{|k|\le N_1(\epsilon)} \max_{t\in [t_0,T]}|\phi_k^m(t)-\phi_k|^2+ \sum_{|k|\le N_1(\epsilon)} \max_{t\in [0,T]}|\phi_k^n(t)-\phi_k|^2\nonumber\\
  &+& 2\left(\sum_{|k|> N_1(\epsilon)} \max_{t\in [t_0,T]}|\phi_k^m(t)-\phi_k|^2\right)^{1/2}\,\left(\sum_{|k|> N_1(\epsilon)} \max_{t\in [t_0,T]}|\phi_k^n(t)-\phi_k|^2\right)^{1/2}\nonumber\\
  &\le&  \frac{\epsilon^2}{2}
  \label{eq:ineq2}
 \end{eqnarray}
 for $m,n >N_2(\epsilon)$. Combining (\ref{eq:sumfractal1}) and (\ref{eq:ineq2}) results in
 \begin{equation}
  \left(\sum_{k} \max_{t\in [t_0,T]}|\phi_k^m(t)-\phi_k^n(t)|^2\right)^{1/2}\le \epsilon,\qquad {\rm for}\,\,\, m,n >N_2(\epsilon).
 \end{equation}
 That is, 
 \begin{equation}
  n(\phi_m-\phi_n)\rightarrow 0 \qquad {\rm as}\,\,\, m,n\rightarrow \infty.\nonumber
 \end{equation}
 
 In conclusion, $\mathcal{A}$ has a finite fractal dimension.

 \ \ \ $\square$

\begin{remark}
 Theorem \ref{theorem:fractal} implies that  strong dissipation renders the dynamics of system  finite-dimensional. Hence, for enhancing dissipation 
a 'freezing' of degrees of freedom (modes) takes place. In particular, the periodic and quasiperiodic breather solution, whose existence was established in Subsections \ref{subsection:periodic} and \ref{subsection:quasiperiodic}, respectively, is composed of a finite number of modes so that the motion is confined to finite-dimensional torus in phase space.
\end{remark}

\subsection{Uniform attractor composed of a single trajectory}\label{section:single}
In this section we show that under strong damping the uniform global attractor is built up from a single trajectory. 
Subsequently, we facilitate this result to prove that a unique periodic and quasiperiodic breather, respectively acts as the uniform global attractor in phase space.

As a first step we demonstrate that under strong damping the system  (\ref{eq:system}) possesses  a single (nonzero) solution contained in a ball in $l^2$ centered at zero.

 We state:
  \begin{proposition}[Existence of a unique solution in $C(\mathbb{R},l^2)$]\label{proposition:uniquesolution1}
  Consider the damped and driven equation  
  \begin{equation}
   i\dot{\phi}_n+\kappa\,[\phi_{n+1}-2\phi_n+\phi_{n-1}]+
 \nu F(|\phi_n|^2)\phi_n+i\gamma\phi_n-g_n=0,\,\,\,n \in \mathbb{Z}.\label{eq:systemphi}
 \end{equation}
 Let 
 \begin{equation}
 \gamma > \left( \frac{1}{\sqrt{2}}\left(1-\frac{1}{\lambda} \right)^{\frac{b-1}{b+1}}\frac{b+1}{b-1}
  \left(\frac{2\sqrt{2}}{a(1+b)} \right)^{2/(1-b)}\parallel g_0\parallel_{C_b(\mathbb{R},l^2)}\right)^{\frac{b-1}{b+1}},\qquad \lambda >1.\label{eq:gammauni}
 \end{equation}
 Then, for   initial data 
   \begin{equation}
    \phi(t_0)=\phi_{t_0}\in l^2,\label{eq:icsuniq}
   \end{equation}
   with 
   \begin{equation}
    ||\phi_{t_0}||_{l^2}
 \le \frac{R}{\lambda},\qquad \lambda >1,
   \end{equation}
 where $R$ is determined by
   \begin{equation}
    R=\min\left\{\left(\frac{\sqrt{\gamma}}{2a}\right)^{1/b}, \left( \frac{2\sqrt{2}\gamma}{a(1+b)} \left(1-\frac{1}{\lambda} \right)\right)^{2/(b-1)}\right\}\label{eq:Rmin}
   \end{equation}
 the equation (\ref{eq:system}) possesses a unique solution $\phi(t)\in B_R\subset C(\mathbb{R},l^2)$.
   \end{proposition}
  
 {\bf Proof:} 
  Taking the spatial Fourier transform of equation (\ref{eq:systemphi})  we obtain
  \begin{equation}
  i\frac{d}{dt}
  \widehat \phi_k +4\kappa \sin ^2\left( \frac{k}{2}\right) \widehat \phi_k
  +i\gamma \widehat \phi_k=\widehat G_k,\label{eq:intuni}
  \end{equation}
  with $G=F(|\phi|^2)\phi+g$.
  
  By DuHamel's principle the solution of (\ref{eq:intuni})  is given by
  \begin{eqnarray}
   \widehat \phi_k(t)=(\Lambda [\widehat \phi_k])(t)
   &=&\exp\left(-\left(4i\kappa \sin ^2\left( \frac{k}{2}\right)+\gamma \right)t\right)\widehat \phi_{0k}\nonumber\\
   &-&i\int_{t_0}^t \exp\left(-\left(4i\kappa \sin ^2\left( \frac{k}{2}\right)+\gamma\right)(t-\tau)\right)
   \widehat G_k(\tau)d\tau.
   \label{eq:intun2}
  \end{eqnarray}
  Proving the existence of a unique solution to the Cauchy problem (\ref{eq:system}),(\ref{eq:icsuniq}) is achieved by  showing that the integral equation (\ref{eq:intun2}) has a (unique) fixed point solution. This is established by verifying that the mapping $\phi \mapsto \Lambda[\phi]$ is a contraction in $B_R\subset C(\mathbb{R},l^2)$ for an appropriate choice of $R>0$. 
  The first step is to  estimate $\Lambda[\Phi]$ in the solution space $C(\mathbb{R},l^2)$.
  Using Minkowski's integral inequality we get
  \begin{eqnarray}
   ||\Lambda [\widehat \phi](t)||_{l^2}&=&\left(\frac{1}{2\pi}\int_0^{2\pi}\left|
   \exp\left(-\left(4i\kappa \sin ^2\left( \frac{k}{2}\right)+\gamma \right)t\right)\widehat \phi_{0k}\right.\right.\nonumber\\
   &-&\left.\left.i \int_{t_0}^t\exp\left(-\left(4i\kappa \sin ^2\left( \frac{k}{2}\right)+\gamma\right)(t-\tau)\right)
   \widehat G_k(\tau)d\tau 
   \right|^2dk\right)^{1/2}\nonumber\\
   &\le& \left(\frac{1}{2\pi}\int_0^{2\pi}\left|
   \exp\left(-\left(4i\kappa \sin ^2\left( \frac{k}{2}\right)+\gamma \right)t\right)\widehat \phi_{0k}\right|^2dk\right)^{1/2}\nonumber\\
   &+&\left(\frac{1}{2\pi}\int_0^{2\pi}\left|\int_{t_0}^t \exp\left(-\left(4i\kappa \sin ^2\left( \frac{k}{2}\right)+\gamma\right)(t-\tau)\right)
   \widehat G_k(\tau)d\tau 
   \right|^2dk\right)^{1/2}\nonumber\\
   &\le& \left( \frac{1}{2\pi}\int_0^{2\pi}\left|\widehat \phi_{0k}\right|^2dk \exp(-2\gamma t)\right)^{1/2}+\left(\frac{1}{2\pi} \int_{t_0}^t \left( \int_0^{2\pi} \left|\widehat G_k(\tau)\right|^2dk\right)^{1/2}\exp(-2\gamma (t-\tau))d\tau\right)^{1/2}
   \nonumber\\
   &\le& ||\phi_0||_{l^2}\exp(-\gamma t)+\left(\int_{t_0}^t ||G(\tau)||_{l^2} \exp(-2\gamma (t-\tau))d\tau\right)^{1/2}.\nonumber
  \end{eqnarray}
  Exploiting that $l^2$ is a Banach algebra, we estimate $||G||_{l^2}$ as follows
  \begin{eqnarray}
   ||G||_{l^2}&=&||F(|\phi|^2)\phi+g\nonumber\\
   &\le& ||F(|\phi|^2)\phi||_{l^2}+||g||_{l^2}\nonumber\\
   &\le& a||\phi||_{l^2}^{1+b}+||g||_{l^2}.
  \end{eqnarray}
  Using that $\sqrt{a+b}\le \sqrt{a}+\sqrt{b}$ we obtain 
  \begin{eqnarray}
   ||\Lambda [\phi]||_{C(\mathbb{R},l^2)}&=& \sup_{t\ge t_0}||\Lambda [\phi](t)||_{l^2}\nonumber\\
   &=&\sup_{t\ge t_0}\left(||\phi_0||_{l^2}\exp(-\gamma (t-t_0))+\left(\int_{t_0}^t ||G(\tau)||_{l^2} \exp(-2\gamma (t-\tau))d\tau\right)^{1/2}\right)\nonumber\\
   &\le& \sup_{t\ge t_0}\left(||\phi_0||_{l^2}\exp(-\gamma (t-t_0))+\sup_{t_0\le \tau \le t}\left[a||\phi(\tau)||_{l^2}^{1+b}+||g(\tau)||_{l^2}\right]^{1/2}\frac{1}{\sqrt{2}\gamma}(1-\exp(-\gamma (t-t_0))\right),\nonumber
  \end{eqnarray}
  where we used the continuous embeddings $l^r\subset l^s,\,\,\,\parallel \phi \parallel_{l^s}\le \parallel \phi\parallel_{l^r},\,\,\,1 \le r\le s \le \infty$.
  We will determine $R$ such that  the map $\phi \mapsto  \Lambda[\phi]$ is  a contraction in the space $C(\mathbb{R},l^2)$.
  Let 
  \begin{equation}
  0<||\phi_0||_{l^2}
 \le \frac{R}{\lambda},\qquad \lambda >1,
  \end{equation}
  and $\phi \in B(0,R) \subset C(\mathbb{R},l^2)$. Then, 
  \begin{equation}
   \frac{1}{\sqrt{2}\gamma}\left( aR^{(1+b)/2}+||g_0||_{C_b(\mathbb{R},l^2)}^{1/2}\right)\le \left(1-\frac{1}{\lambda} \right)R,
  \end{equation}
  is sufficient in order that $\Lambda [\phi] \in B(0,R)$ for all $t> t_0$.
  For a $R_0>0$ for which  
 \begin{equation}
  \frac{1}{\sqrt{2}\gamma}\left( aR_0^{(1+b)/2}+||g_0||_{C_b(\mathbb{R},l^2)}^{1/2}\right)= \left(1-\frac{1}{\lambda} \right)R_0\label{eq:Rnull}
 \end{equation}
 one has that $\Lambda$ maps $B_{R_0}$ into $B_{R_0}$.
 The maximal value of $||g_0||_{C_b(\mathbb{R},l^2)}:=g_M$, for which such a $R_0$ can be found is determined by (\ref{eq:Rnull})
 together with
 \begin{equation}
 \frac{a(1+b)}{2}R_0^{(b-1)/2}=\sqrt{2}\gamma \left(1-\frac{1}{\lambda} \right),\nonumber
 \end{equation}
 resulting in
 \begin{equation}
 R_0=\left( \frac{2\sqrt{2}\gamma}{a(1+b)} \left(1-\frac{1}{\lambda} \right)\right)^{2/(b-1)}.
   \label{eq:Tf}
  \end{equation}
  This yields 
  \begin{equation}
   g_M=\sqrt{2}\gamma ^{\frac{b+1}{b-1}}\left(1-\frac{1}{\lambda} \right)^{\frac{b+1}{b-1}}\left(1-\frac{2}{1+b} 
   \right)
 \left(\frac{2\sqrt{2}}{a(1+b)} \right)^{2/(b-1)},\nonumber
  \end{equation}
  from which follows (\ref{eq:gammauni}).
 
  Further, we need to prove  that  the map $\phi \mapsto  \Lambda[\phi]$ is contractive. That is, there is some $0<k <1$ such that $||\Lambda[\phi]-\Lambda[\psi]||_{C(\mathbb{R},l^2)}=k ||\phi-\psi||_{C(\mathbb{R},l^2)}$ for $\phi,\psi \in B_R$. 
  We have
  \begin{eqnarray}
   ||\Lambda[\phi]-\Lambda[\psi]||_{C(\mathbb{R},l^2)}
   &=&  \sup_{t\ge t_0}||\Lambda[\phi](t)-\Lambda[\psi](t)||_{l^2}\nonumber\\
   &\le &\sup_{t\ge t_0}
   \left(\int_{\mathbb{R}}||G_\phi(\tau) -G_\psi(\tau)||_{l^2}\exp(-\gamma (t-\tau))d\tau\right)^{1/2}\nonumber\\
   &\le& \left(2aR^b\right) \sup_{t \in \mathbb{R}} ||\phi(t)-\psi(t)||_{l^2}\left(\int_{\mathbb{R}}  \exp(-\gamma (t-\tau))d\tau\right)^{1/2}\nonumber\\
   &=&\frac{\left(2aR^b\right)}{\sqrt{\gamma}}||\phi-\psi||_{C(\mathbb{R},l^2)},\nonumber
  \end{eqnarray}
  requiring
  \begin{equation}
   R\le \left(\frac{\sqrt{\gamma}}{2a}\right)^{1/b}.\label{eq:Rf}
  \end{equation}
  Therefore, with  the relations (\ref{eq:Tf}) and 
  (\ref{eq:Rf}) we obtain the claimed result. 
  
  \ \ \ $\square$
  
\vspace*{0.5cm}

\begin{remark}
 For 
 \begin{equation}
  \gamma \ge \max \left\{ \left( 2^{2(1+b)}\,a^2\,||g_0||_{C_b(\mathbb{R},l^2)}^{2b}\right)^{1/(1+2b)},\left( 2^{(1+b)/2}||g_0||_{C_b(\mathbb{R},l^2)}\left( \frac{\sqrt{2}}{a(1+b)}\left(1-\frac{1}{\lambda} \right)\right)^{(b-1)/2}\right)^{(b-1)/(1+b)} \right\} 
 \end{equation}
the absorbing set is contained in the ball radius $R_0$ centered at zero in $l^2$ of . Hence, the absorbing set contains only one solution.
\end{remark}

  Next, we verify that for strong damping  indeed a single trajectory forms  the attractor.
  
   \begin{proposition} 
   Let
  \begin{equation}
  \gamma>\left(a  \parallel g_0\parallel_{C_b(\mathbb{R},l^2)}^b\right)^{1/(1+b)}.\label{eq:gammaunique}
  \end{equation}
  Assume for the initial data of (\ref{eq:system})
  \begin{equation}
   ||\psi_{t_0}||_{l^2}< \frac{\parallel g_0\parallel_{C_b(\mathbb{R},l^2)}}{\gamma}\equiv r_{\mathcal{A}}
  \end{equation}
 
  Then the system  possesses a single solution $\psi(t) \in C(\mathbb{R},l^2)$ with 
  \begin{equation}
   ||\psi||_{l^2}\le  \frac{\parallel g_0\parallel_{C_b(\mathbb{R},l^2)}}{\gamma},
  \end{equation}
  constituting the global uniform attractor of system (\ref{eq:system}).
  \end{proposition}
  
  {\bf Proof:} By Proposition \ref{proposition:globalsolution} and relation (\ref{eq:absset}) solutions of (\ref{eq:system}) on the global uniform attractor satisfy
  \begin{equation}
  || \psi(t)||_{l^2}^2\le \frac{\parallel g_0\parallel_{C_b(\mathbb{R},l^2)}^2}{\gamma},\label{eq:inattrac}
  \end{equation}
  Let $\psi$ and $\phi$ be solutions of (\ref{eq:system}) for initial data  $||\psi_{t_0}||_{l^2}<  r_{\mathcal{A}}$ and $||\phi_{t_0}||_{l^2}<  r_{\mathcal{A}}$, respectively. Since $\psi(t_0),\phi(t_0)\in \mathcal{A}$  we have  
  \begin{equation}
   \parallel\psi(t)\parallel_{l^2}\le \frac{1}{\gamma} \parallel g_0\parallel_{C_b(\mathbb{R},l^2)},\,\,\,\parallel\phi(t)\parallel_{l^2}\le \frac{1}{\gamma} \parallel g_0\parallel_{C_b(\mathbb{R},l^2)},\,\,\,\forall t>t_0.
  \end{equation}
  That is, the corresponding trajectories are contained in the global uniform attractor.
  
  For the distance $||\psi-\phi||_{l^2}$ we derive
  \begin{eqnarray}
    \frac{d}{dt}||\psi-\phi||_{l^2}^2&=& -\gamma\sum_{n}|\psi_n-\phi_n|^2 +i\sum_{n}\left[(F(|\psi_n|^2)\psi_n-F(|\phi_n|^2))\phi_n)\right](\overline{\psi}_n-\overline{\phi}_n)+c.c.\nonumber\\
    &\le& -2(\gamma-a r_{\mathcal{A}}^b)||\psi-\phi||_{l^2}^2,\qquad t\ge t_0,\nonumber
   \end{eqnarray}
   giving 
   \begin{equation}
    ||\psi(t)-\phi(t)||_{l^2}\le \exp[-(\gamma-a r_{\mathcal{A}}^b)\,(t-t_0)]||\psi(t_0)-\phi(t_0)||_{l^2}.\nonumber
   \end{equation}
  Taking into account the relations  (\ref{eq:gammaunique}) and (\ref{eq:inattrac}) and letting $t \rightarrow \infty$ results in
  \begin{equation}
    ||\psi(t)-\phi(t)||_{l^2}\le 0,\nonumber
   \end{equation}
   implying $\psi=\phi$.
  
  Hence,  there exists only a single solution in $C(\mathbb{R},l^2)$ whose associated trajectory 
  coincides with the global uniform attractor completing  the proof. 
   
  \ \ \ $\square$
 \vspace*{0.5cm}
 \begin{corollary}
  If $\gamma \ge a ||g_0||_{C_b(\mathbb{R},l^2)}$, then the single solution coincides with the global uniform attractor.
  In particular,  for strong dissipation, periodic and quasiperiodic forcing exists only a single periodic and quasiperiodic breather solution, 
  respectively that builds up the global uniform attractor. Note that  such a  breather solution is exponentially stable.
 \end{corollary}

%
%
%
%
%

\end{document}